# Phase transitions in sequential weak measurements

Wen-Long Ma[1,3,#], Ping Wang[1,2,3], Weng-Hang Leong[1.3] and Ren-Bao Liu[1,3,*]

1. *Department of Physics, The Chinese University of Hong Kong, Shatin, New Territories, Hong Kong, China*
2. *Beijing Computational Science Research Center, Beijing 100084, China*
3. *Centre for Quantum Coherence, The Chinese University of Hong Kong, Shatin, New Territories, Hong Kong, China*

**Abstract**

Quantum measurements and phase transitions are seemingly uncorrelated topics, but here we show that phase transitions occur in sequential quantum measurements. We find that the probability distribution of the measurement results of a sequence of quantum measurements on a two-level system (e.g. a qubit) is equivalent to the Boltzmann distribution of a classical lattice spin model. So the measurement results present phase transitions similar to those in the lattice spin model. In sequential commuting positive-operator valued measurements, the probability distribution is mapped to a long-range Ising model in the weak-measurement regime, and a projective measurement emerges from a sequence of weak measurement when the strength or the number of measurements becomes above certain critical values, which correspond to a second-order ferromagnetic phase transition of the lattice spin model. These findings not only provide new insights on sequential quantum measurements, but may also have potential applications in quantum technologies.

---

[*] Corresponding author. Email: rbliu@cuhk.edu.hk
[#] Present address: Department of Applied Physics, Yale University, New Haven, Connecticut 06511, USA

**Introduction**

Quantum measurement is a fundamental compound of quantum mechanics [1-5]. It is also important in quantum technologies including quantum computing [6], quantum communication [7] and quantum sensing [8]. The projective measurements are the most commonly considered [6]. More generally the measurement can be the positive-operator valued measurements (POVM) [6], which includes both projective measurements and weak measurements with variable measurement strength. A profound aspect of quantum measurement is that the outputs are randomly distributed. The evolution of the probability amplitudes of the measurement results at time $t$ is governed by the quantum evolution $\exp(-iHt)$ where $H$ is the Hamiltonian. This quantum evolution resembles the random, Boltzmann distribution $\exp(-\beta H)$ of a thermodynamic ensemble at temperature $T = 1/(k_B \beta)$ [9]. Such intriguing similarity between quantum evolution and thermal distribution has stimulated thoughts of unifying the two fundamental laws about stochastics [10-14].

In this work, we discover a surprising connection between quantum mechanics and thermodynamics – the equivalence between sequential measurement outputs and thermodynamic distributions of interacting spin models, and the emergence of projective measurement as a result of phase transitions. For sequential quantum measurements [4, 15, 16], the measurement results obey certain distribution functions depending on the measurement types and on the quantum evolutions between adjacent measurements. We find that for $m$ sequential measurements on, e.g., a two-level system (TLS), the binary results ($\alpha_k = \pm 1$ for $k = 1, \cdots, m$) has a probability distribution $P(\alpha_1, \alpha_2, \cdots, \alpha_m)$ equivalent to the Boltzmann distribution of a classical spin model with each measurement result $\alpha_k$ representing a spin-1/2. Here the number of measurements $m$ can also be understood as the measurement time of a continuous measurement. We focus on two cases where the spin Hamiltonian can be exactly solved: sequential projective measurements can be mapped to the one-dimensional (1D) Ising model with nearest-neighbor interactions [17], and sequential POVM measurement [6] that can be mapped to a long-range Ising model. In the latter case, we find that the projective measurement emerges when the strength or the measurement time are above their respective critical values, which correspond to the second order ferromagnetic phase transition in the long-range Ising model [18].

## Statistics of sequential quantum measurements

For *m* successive POVM measurements on a TLS (e.g. a spin-1/2 qubit), the probability to obtain the measurement result $(\alpha_1, \alpha_2, \cdots, \alpha_m)$ is

$$P(\alpha_1, \alpha_2, \cdots, \alpha_m) = \text{Tr}\left[M_{\alpha_m} \cdots M_{\alpha_2} M_{\alpha_1} |\psi_0\rangle\langle\psi_0| M_{\alpha_1}^\dagger M_{\alpha_2}^\dagger \cdots M_{\alpha_m}^\dagger \right], \qquad (1)$$

where $|\psi_0\rangle$ is the initial state of the TLS and $\{M_{\alpha_k}\}$ ($\alpha_k = \pm 1$) are the set of POVM operators for the *k*th measurement satisfying $\sum_{\alpha_k} M_{\alpha_k}^\dagger M_{\alpha_k} = I$. Here we assume that the evolution of the TLS between measurements has been absorbed into the POVM operators.

Below we will show how to map measurement statistics in Eq. (1) to the occupation probability of the configuration $(\alpha_1, \alpha_2, \cdots, \alpha_m)$ of a classical lattice spin model, i.e.

$$P(\alpha_1, \alpha_2, \cdots, \alpha_m) = \exp\left[-H(\alpha_1, \alpha_2, \cdots, \alpha_m)\right], \qquad (2)$$

where $H(\alpha_1, \alpha_2, \cdots, \alpha_m)$ is the Hamiltonian of the lattice spin model with $\alpha_k$ denoting the *k*th lattice spin (temperature absorbed into the Hamiltonian). Obviously, the Hamiltonian can be written as $H(\alpha_1, \alpha_2, \cdots, \alpha_m) = -\ln\left[P(\alpha_1, \alpha_2, \cdots, \alpha_m)\right]$. Below we focus on two cases where exact solutions are possible.

## Case I: sequential projective measurements

For *m* successive projective measurements on the TLS, the projective operators are

$$M_{\alpha_k} = \frac{1}{2}\left[I + \alpha_k (\hat{\boldsymbol{\sigma}} \cdot \boldsymbol{n}_k)\right] = |\alpha_k\rangle\langle\alpha_k|, \qquad (3)$$

where $\alpha_k = \pm 1$ represents the binary measurement results, $\hat{\boldsymbol{\sigma}} = (\sigma_x, \sigma_y, \sigma_z)$ are the Pauli matrices of the TLS and $n_k$ is the unit vector of the $k$th measurement axis, and $|\alpha_k\rangle$ is the eigenstate of $\hat{\sigma} \cdot n_k$ with eigenvalue $\alpha_k$. Suppose the initial state of the TLS is $|\psi_0\rangle = |\alpha_0\rangle$ with $\alpha_0 = \pm 1$, then the probability distribution of the measurement results can be directly calculated as [4]

$$P_I(\alpha_1, \alpha_2, \cdots, \alpha_m) = \frac{1}{2^m} \prod_{k=1}^{m} \left[1 + \cos(\phi_{k-1,k}) \alpha_{k-1} \alpha_k\right], \quad (4)$$

where $\phi_{k-1,k} = \arccos(n_{k-1} \cdot n_k) \in [0, \pi]$ denotes the angle between the $(k-1)$th and $k$th measurement axes.

The probability distribution for sequential projective measurements [Eq. (4)] is exactly the normalized partition function of a classical 1D Ising model with nearest-neighbour interaction,

$$H_I(\alpha_1, \alpha_2, \cdots, \alpha_m) = -\sum_{k=1}^{m} J_{k-1,k} \alpha_{k-1} \alpha_k, \quad (5)$$

where $J_{k-1,k} = \tanh^{-1}\left[\cos(\phi_{k-1,k})\right]$ is the coupling strength between two neighbouring spins. If $\phi_{k-1,k} \in [0, \pi/2)$, $J_{k-1,k} > 0$ corresponding to a ferromagnetic coupling; if $\phi_{k-1,k} \in (\pi/2, \pi]$, $J_{k-1,k} < 0$ corresponding to an anti-ferromagnetic coupling; if $\phi_{k-1,k} = \pi/2$, $J_{k-1,k} = 0$ corresponding to the non-interacting case.

The correlation function between the results of the $j$th measurement and the $(j+n)$th measurement, the same as the correlation function of the 1D Ising model in Eq. (5), can be obtained as

$$\langle \alpha_j \alpha_{j+n} \rangle = \prod_{k=j+1}^{j+n} \tanh(J_{k-1,k}) = \prod_{k=j+1}^{j+n} \cos(\phi_{k-1,k}). \quad (6)$$

Let us consider the specific case where $J_{0,1} = J_{1,2} \cdots = J_{m-1,m} = J$ and $\phi_{0,1} = \phi_{1,2} \cdots = \phi_{m-1,m} = \phi$. For the ferromagnetic coupling ($J > 0$, i.e. $0 < \cos\phi < 1$), the correlation function $\langle \alpha_j \alpha_{j+n} \rangle = \cos^n(\phi)$ shows a power-law decay with respect to the distance between the two lattice spins, indicating a paramagnetic phase. If $\cos\phi = 1$, the correlation function is a constant ($\langle \alpha_j \alpha_{j+n} \rangle = 1$), indicating a ferromagnetic phase transition for infinite coupling ($J = +\infty$) or zero temperature. This phase transition can be intuitively understood: The condition $\cos\phi=1$ corresponds to the case that all the projective measurements are along the same axis, therefore the first measurement collapses the TLS into an eigenstate of the projective operator and all the subsequent measurements will give the same results. The case for the anti-ferromagnetic coupling can be similarly analysed, except that at the anti-ferromagnetic phase transition point ($\cos\phi = -1$ and $J = -\infty$), the correlation function becomes $\langle \alpha_j \alpha_{j+n} \rangle = (-1)^n$.

## Case II: sequential commuting POVM measurements

Now we consider $m$ successive commuting POVM measurements on the TLS with the POVM operators defined as [19]

$$M_{\alpha_k} = \frac{1}{\sqrt{2}} \left[ \cos(\theta_k) I + \sin(\theta_k) \alpha_k \sigma_z \right], \qquad (7)$$

where $\theta_k \in [0, \pi/4]$. The measurement strength $\lambda_k$ is defined by $\lambda_k = \sin^2(2\theta_k)$. When the measurement strength increases from 0 to 1, the $k$th measurement continuously changes from weak measurement to strong projective measurement.

Suppose the initial state of the TLS is $|\psi_0\rangle = C_0^+ |+1\rangle + C_0^- |-1\rangle$ with $|\pm 1\rangle$ being the eigenstates of $\sigma_z$ and $|C_0^+|^2 + |C_0^-|^2 = 1$. The unnormalized state of the TLS after $m$ measurements is

$$|\psi_m\rangle = M_{\alpha_m} \cdots M_{\alpha_2} M_{\alpha_1} |\psi_0\rangle = C_m^+ |+1\rangle + C_m^- |-1\rangle, \qquad (8)$$

with

$$C_m^\pm = \frac{1}{\sqrt{2}} \left[ \cos(\theta_m) \pm \sin(\theta_m) \alpha_m \right] C_{m-1}^\pm = \frac{C_0^\pm}{2^{m/2}} \prod_{k=1}^m \left[ \cos(\theta_k) \pm \sin(\theta_k) \alpha_k \right], \qquad (9)$$

and the normalized state is $|\psi'_m\rangle = |\psi_m\rangle/\sqrt{\langle\psi_m|\psi_m\rangle}$. Denote the Bloch vector components of the final state as $r^i_m = \langle\sigma_i\rangle = \langle\psi'_m|\sigma_i|\psi'_m\rangle$ ($i = x, y, z$) ($r_m = 1$ for a pure state), the probability distribution for the measurement results is analytically derived as

$$P_{\mathrm{II}}(\alpha_1, \alpha_2, \cdots, \alpha_m) = \langle\psi_m|\psi_m\rangle = \frac{1}{2^{m+1}}\left[(1+r^z_0)\prod_{k=1}^{m}(1+\sqrt{\lambda_k}\alpha_k) + (1-r^z_0)\prod_{k=1}^{m}(1-\sqrt{\lambda_k}\alpha_k)\right], \quad (10)$$

where $r^z_0 = |C^+_0|^2 - |C^-_0|^2$ is the $z$-component of the Bloch vector of the initial state.

The lattice spin Hamiltonian corresponding to the probability distribution of the sequential POVM measurement is

$$H_{\mathrm{II}}(\alpha_1, \alpha_2, \cdots, \alpha_m) = -\ln\left\{\frac{1}{2^{m+1}}\left[(1+r^z_0)\prod_{k=1}^{m}(1+\sqrt{\lambda}\alpha_k) + (1-r^z_0)\prod_{k=1}^{m}(1-\sqrt{\lambda}\alpha_k)\right]\right\}, \quad (11)$$

where we have assumed that all the sequential POVM measurements in Eq. (7) are the same with $\lambda_1 = \lambda_2 = \cdots = \lambda_m = \lambda$. We identify the order parameter of the above spin model as the measurement polarization $X = q/m - 1/2$ with $q$ being the number of measurements with result $\alpha = +1$, then the probability distribution of $X$ is

$$P(X) = \left[(1-\lambda)/4\right]^{m/2} C_m^{m(X+1/2)}\left[\cosh(\ln(\eta)mX) + r^z_0\sinh(\ln(\eta)mX)\right], \quad (12)$$

where $\eta = (1+\sqrt{\lambda})/(1-\sqrt{\lambda})$. We define the free energy as

$$F(X) = -\ln[P(X)] \approx m\varphi(X) - \ln\left[\cosh(\ln(\eta)mX) + r^z_0\sinh(\ln(\eta)mX)\right], \quad (13)$$

where $\varphi(X) = (1/2 + X)\ln(1/2 + X) + (1/2 - X)\ln(1/2 - X)$ [20]. In $F(X)$, the first and the second parts represent the entropy and the internal energy of the lattice spin model, respectively. The free energy takes the minimum when

$$\frac{\partial F(X)}{\partial X} = m\left\{\ln\left(\frac{1+2X}{1-2X}\right) - \ln(\eta)\frac{\sinh(\ln(\eta)mX) + r^z_0\cosh(\ln(\eta)mX)}{\cosh(\ln(\eta)mX) + r^z_0\sinh(\ln(\eta)mX)}\right\} = 0. \quad (14)$$

After solving the above equation for $X$, the z-component of the Bloch vector of the TLS after $m$ POVM measurements can be obtained as

$$r_m^z = \frac{\sinh(\ln(\eta)mX) + r_0^z \cosh(\ln(\eta)mX)}{\cosh(\ln(\eta)mX) + r_0^z \sinh(\ln(\eta)mX)}, \quad (15)$$

and the x,y-components are $r_m^{x/y} = r_0^{x/y} \big/ \left[\cosh(\ln(\eta)mX) + r_0^z \sinh(\ln(\eta)mX)\right]$.

To simplify the discussion we consider the case that the initial states of the TLS lies in the equatorial plane of the Bloch sphere with $r_0^z = 0$. In this case the probability distribution becomes

$$P_{\text{II}}^B(\alpha_1, \alpha_2, \cdots, \alpha_m) = \frac{1}{2^{m+1}}\left[\prod_{k=1}^{m}(1+\sqrt{\lambda}\alpha_k) + \prod_{k=1}^{m}(1-\sqrt{\lambda}\alpha_k)\right], \quad (16)$$

and the corresponding 1D lattice spin Hamiltonian is

$$H_{\text{II}}^B(\alpha_1, \alpha_2, \cdots, \alpha_m) = -\ln\left\{\frac{1}{2}\left[\prod_{k=1}^{m}(1+\sqrt{\lambda}\alpha_k) + \prod_{k=1}^{m}(1-\sqrt{\lambda}\alpha_k)\right]\right\}, \quad (17)$$

where we have dropped the constant $m$ in $H_{\text{II}}$. In the weak-measurement regime ($\lambda \ll 1$), the spin Hamiltonian in Eq. (17) is equivalent to the long-range ferromagnetic Ising model up to leading-order terms ($\propto \lambda$),

$$H_{\text{II}}^B(\alpha_1, \alpha_2, \cdots, \alpha_m) \approx -\lambda \sum_{j<k}^{m} \alpha_j \alpha_k. \quad (18)$$

The free energy of the lattice spin model becomes

$$F(X) = m\varphi(X) - \ln\left[\cosh(\ln(\eta)mX)\right], \quad (19)$$

which depends on both the measurement strength $\lambda$ and the measurement time $m$. For a fixed $m$, $F(X)$ shows spontaneous symmetry breaking as $\lambda$ is increased [Fig. 1(a)]; Similarly, for a fixed $m$, $F(X)$ also shows spontaneous symmetry breaking as $m$ is increased [Fig. 1(a)]. This shows a phase transition between the unpolarized phase and the polarized phase in the two-dimensional parameter space $(\lambda, m)$, which is verified by a Monte Carlo simulation of $10^4$ samples of sequential POVM measurements (Fig. 2).

The distance between the two valleys in the polarized phase increases with the measurement strength $\lambda$ but is independent of the number of measurements. The free energy takes the minimum when

$$\frac{\partial F(X)}{\partial X} = m\left\{\ln\left(\frac{1+2X}{1-2X}\right) - \ln(\eta)\tanh\left[\ln(\eta)mX\right]\right\} = 0, \tag{20}$$

By solving the above equation, we find that if $m\ln^2(\eta) < 4$, the free energy has only one minimum at $X = 0$ corresponding to the unpolarized phase, while if $m\ln^2(\eta) > 4$, the free energy has two minima located at $(-1/2, 0)$ and $(1/2, 0)$ corresponding to the polarized phase. So the phase transition occurs when the measurement times and the measurement strength satisfy $m\ln^2(\eta) = 4$ [Fig. 3(a)]. For weak measurement ($\lambda \ll 1$), the phase boundary is $m\lambda = 1$, which coincides with that for the approximated long-range Ising model in Eq. (18) [Fig. 3(a)].

However, the order parameters as functions of the measurement time and measurement strength are quite different for the exact spin model [Eq. (17)] and the approximated long-range Ising model [Eq. (18)], as shown in Fig. 3(b)(c). In the exact model, for fixed measurement times $m$, the order parameter $X$ quickly increases above the critical measurement strength $\lambda_c = \tanh^2(m^{-1/2})$ and then increases linearly with $\lambda$ as $X = \pm\sqrt{\lambda}/2$; for a fixed $\lambda$, $X$ also quickly increases above the critical measurement time $m_c = 4/\ln^2\left[(1+\sqrt{\lambda})/(1-\sqrt{\lambda})\right]$ and approaches the constant $X = \pm\sqrt{\lambda}/2$ as $m$ is further increased. This implies that in the polarized phase the measurement polarization is proportional to the measurement strength but independent of the measurement time. Moreover, the derivative of the order parameter as a function of the measurement strength or measurement time shows a finite jump at the critical points [Fig. 3(d)(e)], which is a signature of second-order phase transitions. However, for the long-range ferromagnetic Ising model, it is $m\lambda$ that influences the ferromagnetic phase transition, and the polarization reaches the maximum value $X = \pm 1/2$ in the ferromagnetic phase.

Moreover, the final state polarization of the TLS also shows a phase-transition behavior depending on the measurement time and measurement strength. For a fixed measurement time $m$, the final state polarization keeps almost unchanged compared to the initial one with the measurement strength below the critical value $\lambda_c$ but quickly becomes fully polarized to the north or south pole as $\lambda$ increases above $\lambda_c$ [Fig. 4(d)]. Similar behavior is observed for a fixed measurement strength and increasing measurement time [Fig. 4(e)]. When the state polarization begins, the TLS has the same probability to be polarized to the north or south pole, and it has to decide which path to choose. This is quite similar to the spontaneous symmetry breaking in statistical physics.

If the initial state of the TLS is in the north or south pole with $r_0^z = \pm 1$, then the probability distribution becomes

$$P_{\text{II}}^A(\alpha_1, \alpha_2, \cdots, \alpha_m) = \frac{1}{2^m} \prod_{k=1}^{m} \left(1 + \sqrt{\lambda}\alpha_0 \alpha_k \right), \tag{21}$$

with $\alpha_0 = \left|r_0^z\right|/r_0^z$. This is just the normalized probability of the configuration $(\alpha_1, \alpha_2, \cdots, \alpha_m)$ for $m$ independent paramagnetic classical spins with the Hamiltonian

$$H_{\text{II}}^A(\alpha_1, \alpha_2, \cdots, \alpha_m) = -\omega \alpha_0 \sum_{k=1}^{m} \alpha_k. \tag{22}$$

where $\omega = \tanh^{-1}\left(\sqrt{\lambda}\right)$ is the effective energy of the spins and $\alpha_0$ determines the magnetic field direction. So the measurement polarization $X$ can be understood as the average magnetic polarization of all the spins, i.e. $X = \tanh(\omega)/2 = \sqrt{\lambda}/2$, where the free energy of the spin model has the minimum [Fig. 4(a)]. The reason is that the state of TLS is unchanged by the measurements, as can be seen from Eq. (15) and Fig. 4(d)(e), so the probability to obtain different results are the same for all the measurements. In this case, there is no phase transition.

If the initial states of the TLS are anywhere on the Bloch sphere other than the north or south poles or the equator with $|r_0^z| \in (0,1)$, in the weak-measurement regime ($\lambda \ll 1$), the probability distribution in Eq. (10) can be mapped to the long-range ferromagnetic Ising model under an external magnetic field up to the leading-order terms ($\propto \lambda$),

$$H_{\text{II}}^C(\alpha_1, \alpha_2, \cdots, \alpha_m) \approx -r_0^z \sqrt{\lambda} \sum_{k=1}^{m} \alpha_k - \lambda \sum_{j<k}^{m} \alpha_j \alpha_k, \tag{23}$$

where the magnetic field is proportional to the z-component of Bloch vector polarization of the initial state. In this case, the free energy becomes unsymmetrical in the polarized phase and therefore the measurement polarization has a preferred value, i.e. $X = \sqrt{\lambda}/2$ ($X = -\sqrt{\lambda}/2$) for $r_0^z > 0$ ($r_0^z < 0$), and the probability in the preferred value is about $(1+|r_0^z|)/(1-|r_0^z|)$ times that in the unpreferred value. The measurement polarization $X$ as a function of measurement time $m$ and measurement strength $\lambda$ changes more and more smoothly as $|r_0^z|$ increases and the phase-transition behaviours gradually disappear [Fig. 4(b)(c)]. Moreover, the final state of the TLS is also gradually polarized toward the north (south) pole for $r_0^z > 0$ ($r_0^z < 0$) as the measurement times or measurement strength increases [Fig. 4(d)(e)].

## Example- nuclear spin polarization by an ancillary electron spin

As an example, let us consider an electron spin (e.g. two energy levels of a nitrogen-vacancy electron spin) and a nuclear spin (e.g. a $^{13}$C nuclear spin in diamond). The POVM measurement of the nuclear spin in Eq. (7) can be realized by coupling it to the electron spin and then performing projective measurements on the electron spin [21, 22]. The Hamiltonian of the electron spin ($S = 1/2$) and the nuclear spin ($I = 1/2$) is

$$H = A S_z I_z + \omega I_z, \tag{24}$$

where $S_z$ ($I_z$) is the electron (nuclear) spin operator with eigenstates $|\pm\rangle_e$ ($|\pm\rangle_n$), $A$ is the coupling strength and $\omega$ is the Larmor frequency of the nuclear spin. The target spin evolution operator conditioned on the sensor spin state is $U_n^{(\pm)}(t) = e^{-i(\omega \pm A/2)I_z t}$. We apply the Ramsey sequence [21] to the electron spin with the propagator of the whole system as

$$U(t) = R_e^x(\pi/2)\left(U_n^{(+)}(t)|+\rangle_e{}_e\langle+| + U_n^{(-)}(t)|-\rangle_e{}_e\langle-|\right)R_e^y(\pi/2), \quad (25)$$

where $R_e^j(\pi/2) = e^{-i\pi S_j/2}$ $(j = x, y)$ denotes the $\pi/2$ pulse for the electron spin along different axes. Suppose the initial state of the whole system is $|+\rangle_e \otimes |\psi_0\rangle$ with $|\psi_0\rangle = C_0^+|+\rangle_n + C_0^-|-\rangle_n$ denoting the initial target spin state, then projective measurements on the sensor spin with $M_e^{(\alpha)} = (I + 2\alpha S_z)/2$ $(\alpha = \pm 1)$ is equivalent to a POVM measurement on the nuclear spin, i.e.

$$M_n^{(\alpha)}|\psi_0\rangle\langle\psi_0|\left(M_n^{(\alpha)}\right)^\dagger = \text{Tr}_e\left[M_e^{(\alpha)}U(t)\left(|+\rangle_e{}_e\langle+| \otimes |\psi_0\rangle\langle\psi_0|\right)U^\dagger(t)\left(M_e^{(\alpha)}\right)^\dagger\right], \quad (26)$$

where $M_n^{(\alpha)} = \left(U_n^{(+)} + i\alpha U_n^{(-)}\right)/2 = e^{-i(\omega I_z t - \pi/4)}\left[\cos(\theta)I - 2\alpha\sin(\theta)I_z\right]/\sqrt{2}$ with $\theta = At/2$. Note that $M_n^{(\alpha)}$ is the same POVM operator as that in Eq. (8) except that there is an additional evolution operator $e^{-i(\omega I_z t - \pi/4)}$ which is independent of the measurement results and has no effect on the probability distribution. By repetitively applying the Ramsey sequence to the electron spin, sequential POVM measurements are performed on the nuclear spin with the measurement strength tuned by the time delay $t$ between the two $\pi/2$ pulses [21], and the nuclear spin is polarized to $|+\rangle_n$ ($|-\rangle_n$) with the probability equal to the probability amplitude of the initial state $|C_0^+|^2$ ($|C_0^-|^2$). After spontaneous symmetry breaking at $m = m_c$, the nuclear spin will be trapped in the polarized state by the sequential weak measurement.

## Conclusions

We establish the connections between the probability distribution of sequential quantum measurement on a TLS and the statistical mechanics of 1D spin models. Therefore, the statistics and phase transitions of the spin chains can be effectively simulated by measuring a single qubit. For sequential projective measurements, the measurement results effectively simulate the 1D Ising model with nearest-neighbour interactions; for sequential commuting POVM measurements, the measurement is mapped to ferromagnetic long-range Ising models. We find a polarized-to-unpolarized phase transition in the sequential POVM measurements dependent on the measurement time and measurement strength.

## Acknowledgements

This work was supported by Hong Kong Research Grants Council.

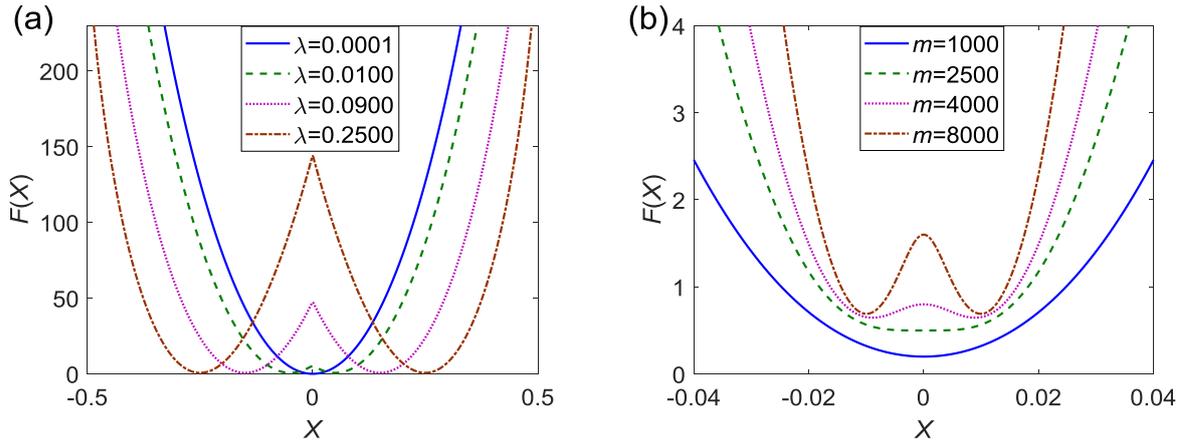

**Fig. 1.** Free energy landscape of the spin model corresponding to a sequential POVM measurement. (a) Free energy as a function of measurement polarization $X$ for different measurement strength $\lambda$ with the measurement time fixed at $m=1000$. (b) Free energy as a function of measurement polarization $X$ for different measurement times $m$ with the measurement strength fixed at $\lambda=0.0004$. Here the initial state of the TLS is in the equator of the Bloch sphere with $r_0^z=0$.

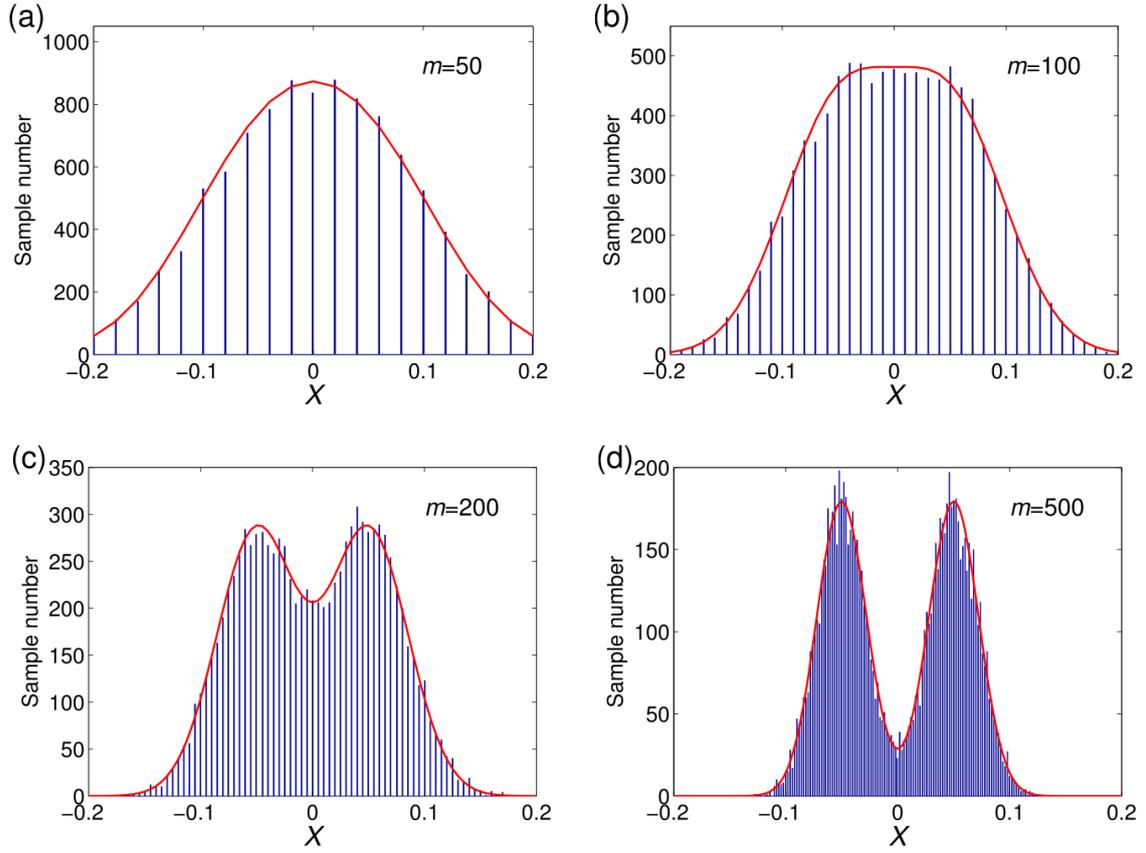

**Fig. 2.** The histogram of the number of samples with respect to the measurement polarization $X$ in the Monte Carlo simulation of sequential POVM measurements in Model (II) for different measurement time: (a) $m=50$, (b) $m=100$, (c) $m=200$ and (d) $m=500$. The red solid lines represent the exact probability distribution in Eq. (16). The measurement strength $\lambda=0.01$. The Monte Carlo simulation contains $10^4$ samples of sequential measurements from the same initial state ($r_0^z=0$).

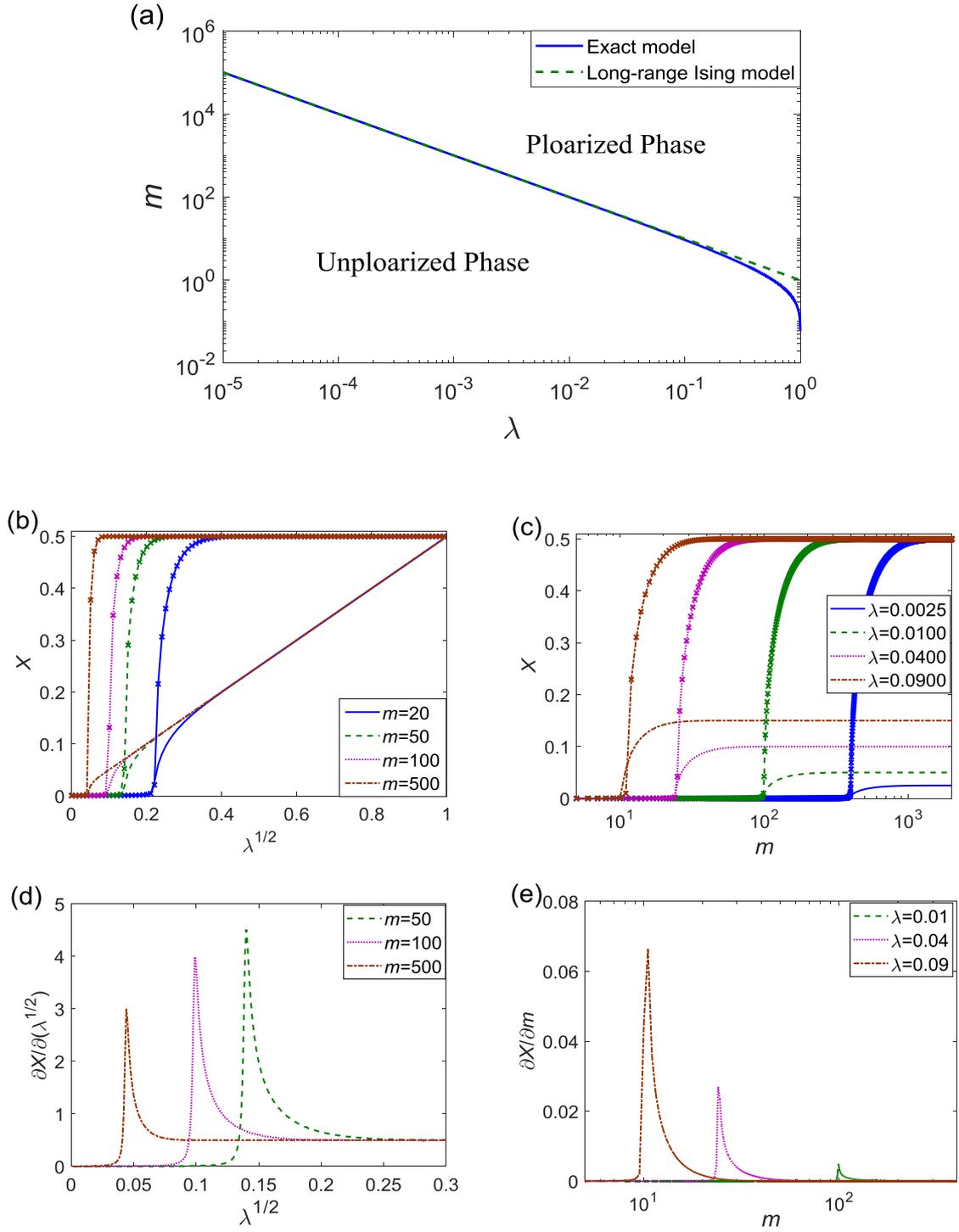

**Fig. 3.** Phase transitions of sequential weak measurement. (a) Phase-transition measurement time $m_c$ as a function of the measurement strength $\lambda$. (b), (d) The measurement polarization $X$ (solid lines) and $\partial X/\partial\sqrt{\lambda}$ as functions of the square root of the measurement strength for different measurement times. (c), (e) The measurement polarization $X$ and $\partial X/\partial\sqrt{\lambda}$ as functions of the measurement times

for different measurement strengths. The lines without (with) crosses represent the results from the exact model (the approximate long-range Ising model). The initial state is in the equator of the Bloch sphere ($r_0^z = 0$).

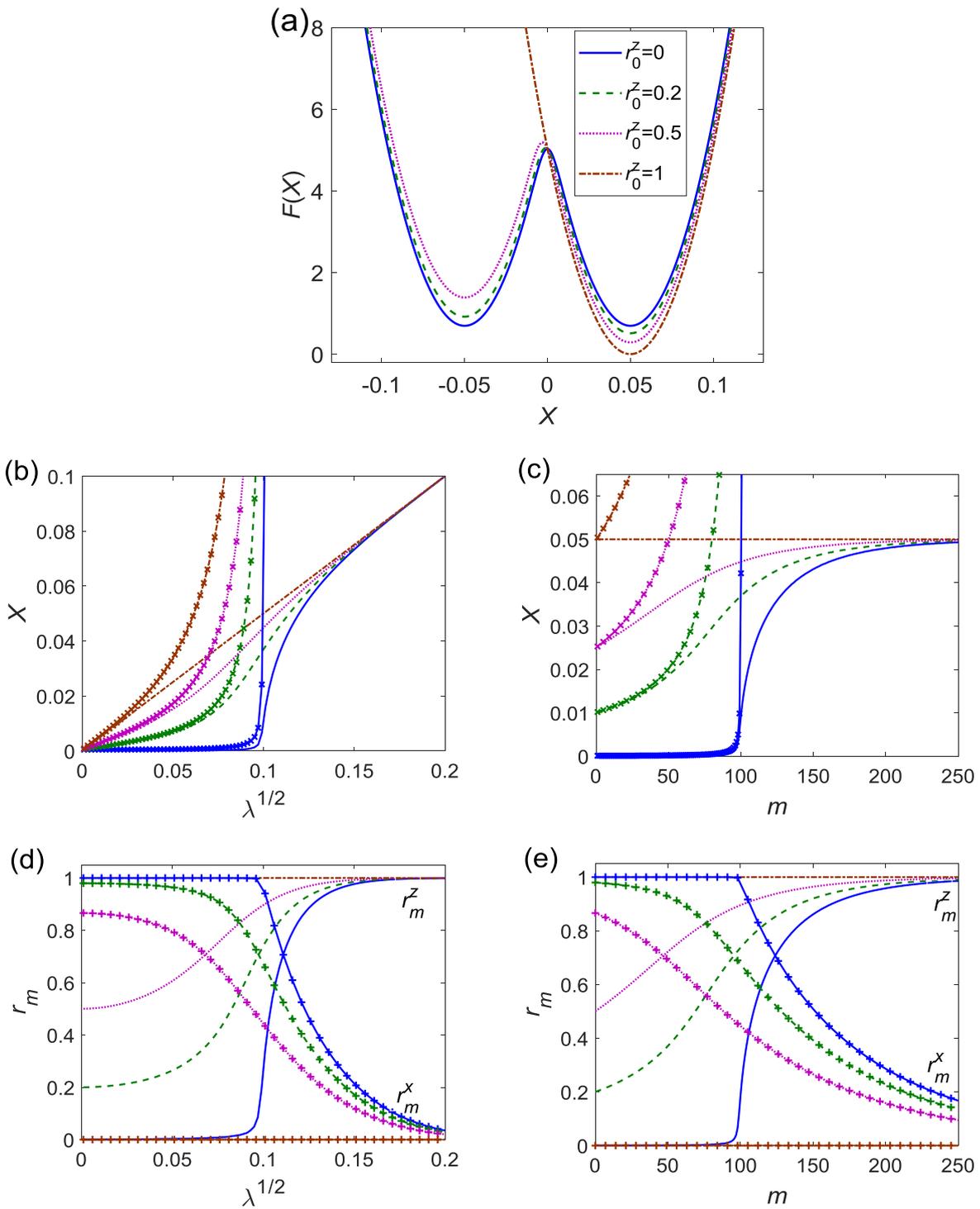

**Fig. 4.** Effects of initial states on the phase transitions of sequential weak measurement. Different initial states (given by the initial polarization $r_0^z$) are represented by different line colors. (a) Free energy as a function of the measurement polarization $X$. The measurement strength and measurement time are $m = 1000$ and $\lambda = 0.01$, respectively.

(b), (d) The measurement polarization $X$ and the final Bloch vector polarization $r_m$ as functions of the square root of the measurement strength with measurement times fixed at $m = 100$. (c), (e) The measurement polarization $X$ and the final Bloch vector polarization $r_m$ as functions of the measurement time with the measurement strength fixed at $\lambda = 0.01$. In (b) and (c), the lines without (with) crosses represent the results from the exact model (the approximate long-range Ising model). In (d) and (e), the lines without (with) plus signs represent the $z$ ($x$) component of the final Bloch vector. Initially the TLS is in a pure state with $r_0 = 1$ and $r_0^y = 0$.